\documentclass{optica-article}

\journal{opticajournal} 

\articletype{Research Article}

\newcommand{\LB}{\textit{LiteBIRD}}

\begin{document}

\title{Modified crossed Dragone optical design of the LiteBIRD low-frequency telescope}

\author{Frederick Matsuda,\authormark{1,*}, Shugo Oguri,\authormark{1}, Yutaro Sekimoto\authormark{1}, Aritoki Suzuki\authormark{2}, Hayato Takakura\authormark{1}, Shingo Kashima\authormark{3}}

\address{\authormark{1}Institute of Space and Astronautical Science (ISAS), Japan Aerospace Exploration Agency (JAXA), Sagamihara, Kanagawa 252-5210, Japan}

\address{\authormark{2}Lawrence Berkeley National Laboratory (LBNL), Computational Cosmology Center, Berkeley, CA 94720, USA}

\address{\authormark{3}National Astronomical Observatory of Japan, Mitaka-shi, Tokyo, Japan}

\email{\authormark{*}matsuda.frederick@jaxa.jp} 


{\noindent \copyright 2025 Optica Publishing Group. One print or electronic copy may be made for personal use only. Systematic reproduction and distribution, duplication of any material in this paper for a fee or for commercial purposes, or modifications of the content of this paper are prohibited.}

{\noindent This article is published in Applied Optics Vol. 64, Issue 14, pp. 4050-4060 (2025), and may be found at https://doi.org/10.1364/AO.551522.}

\begin{abstract*} 

 \LB\ is a JAXA-led international project aimed at measuring the cosmic microwave background (CMB) polarization with high sensitivity to detect polarization $B$ modes. This detection would provide evidence of inflation.
 \LB\ will observe the full sky for three years at the L2 Lagrange point of the Earth-Sun system across 34--448\,GHz, and is expected to launch in the Japanese fiscal year of 2032. The Low-Frequency Telescope (LFT) will observe in the 34--161\,GHz range implementing a modified crossed Dragone (MCD) reflective optical design optimized for high optical performance across a wide $18^{\circ} \times 9^{\circ}$ field-of-view (FOV). In this paper, we report the LFT optical design details including its optimization and optical performance assessed using optical simulations. The MCD design consists of a paraboloidal primary and a hyperboloidal secondary reflector with polynomial correction terms up to 7th order, achieving Strehl ratios $\ge 0.97$ at 161\,GHz across the FOV. The Mueller QU (UQ) cross-polarization response is $\le -26.9$ dB at 34\,GHz. The simulated beam sizes are $< 78^{\prime}$ at 34\,GHz. The simulated sidelobe response for the direct and diffuse triple reflection sidelobes are estimated to be $< -57$ dB, and for the focused triple reflection sidelobe $< -37$ dB at 34\,GHz. The LFT optical design satisfies all the optical requirements and specifications for the project, and is compatible with the \LB\ science goals.

\end{abstract*}

\section{Introduction}\label{sec:intro}

The cosmic microwave background (CMB) contains rich information that allows us to probe the physics of the early universe. If inflation occurred in the early universe, inflationary primordial gravitation waves are theorized to imprint a distinct large angular scale CMB polarization signal called the primordial $B$ modes. While the temperature (unpolarized) component of the CMB has been measured and studied in detail, the search of $B$ modes continues with high-precision CMB polarization measurements\cite{Kamionkowski_2016}. 

The \LB\ mission is a JAXA-led international project aimed at high-sensitivity measurements of the primordial $B$ modes and that would provide evidence of inflation. \LB\ will observe the full sky for three years at the L2 Lagrangian point of the Earth-Sun system across a wide frequency range of 34--448 GHz and constrain the tensor-to-scalar ratio parameter to $\delta(r) < 0.001$ sensitivity\cite{LB_ptep_2022}. The spacecraft payload module (PLM) consists of two instruments: the Low-Frequency Telescope (LFT) which will observe at 34--161\,GHz, and the Mid and High-Frequency Telescopes (MHFT) which will observe at 89--448\,GHz. \LB\ is expected to launch in the Japanese fiscal year of 2032 using the JAXA H3 rocket. 

The LFT is a modified crossed Dragone reflective telescope that has been optimized to achieve high optical performance levels and large throughput required for high-sensitivity CMB polarization observations from space. In this paper, we describe the details of the LFT optical design including its optimization, and evaluate its optical performance using geometric and physical optics simulation calculations. We compare the design performance levels to the \LB\ optical requirements derived from the science goals. We show that the LFT optical design satisfies the requirements and is suitable for the \LB\ science. In Section \ref{sec:lft}, we provide an overview of the LFT, and in Section \ref{sec:req}, we summarize the optical requirements. In Section \ref{sec:opt-design}, we describe the optical design geometry and performances including Strehl ratio, telecentricity, F/\#, polarization angle rotation, and cross-polarization. In Section \ref{sec:po-perf}, we estimate the main beam and sidelobe performances. 

\begin{figure}[htb]
    \centering
    \includegraphics[width=0.65\hsize,pagebox=cropbox,clip]{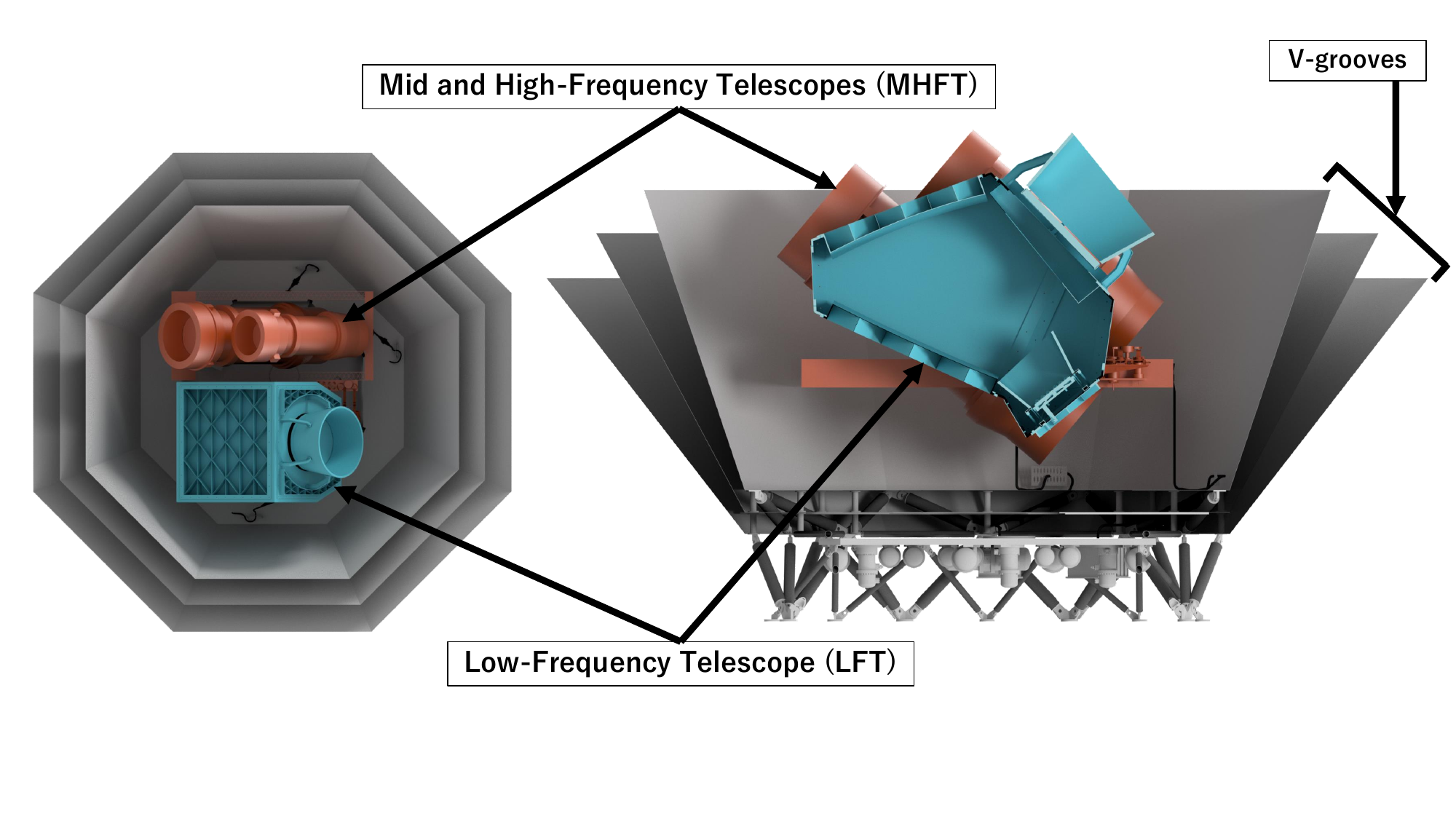}
    \caption{PLM design including the LFT, MHFT, and surrounding V-groove structures.}
    \label{fig:plm-overview}
\end{figure}

The MHFT consists of two refractive telescopes that will observe from 89--224\,GHz and 166--448\,GHz. The optical and mechanical design of the MHFT is described in Montier et al., 2020\cite{Montier_2020}. The LFT and MHFT within the PLM is shown in Figure \ref{fig:plm-overview}. This paper will focus only on the LFT optical design.

\section{\LB\ Low-Frequency Telescope}\label{sec:lft}

\begin{figure}[htb]
    \centering
    \includegraphics[width=0.8\hsize,pagebox=cropbox,clip]{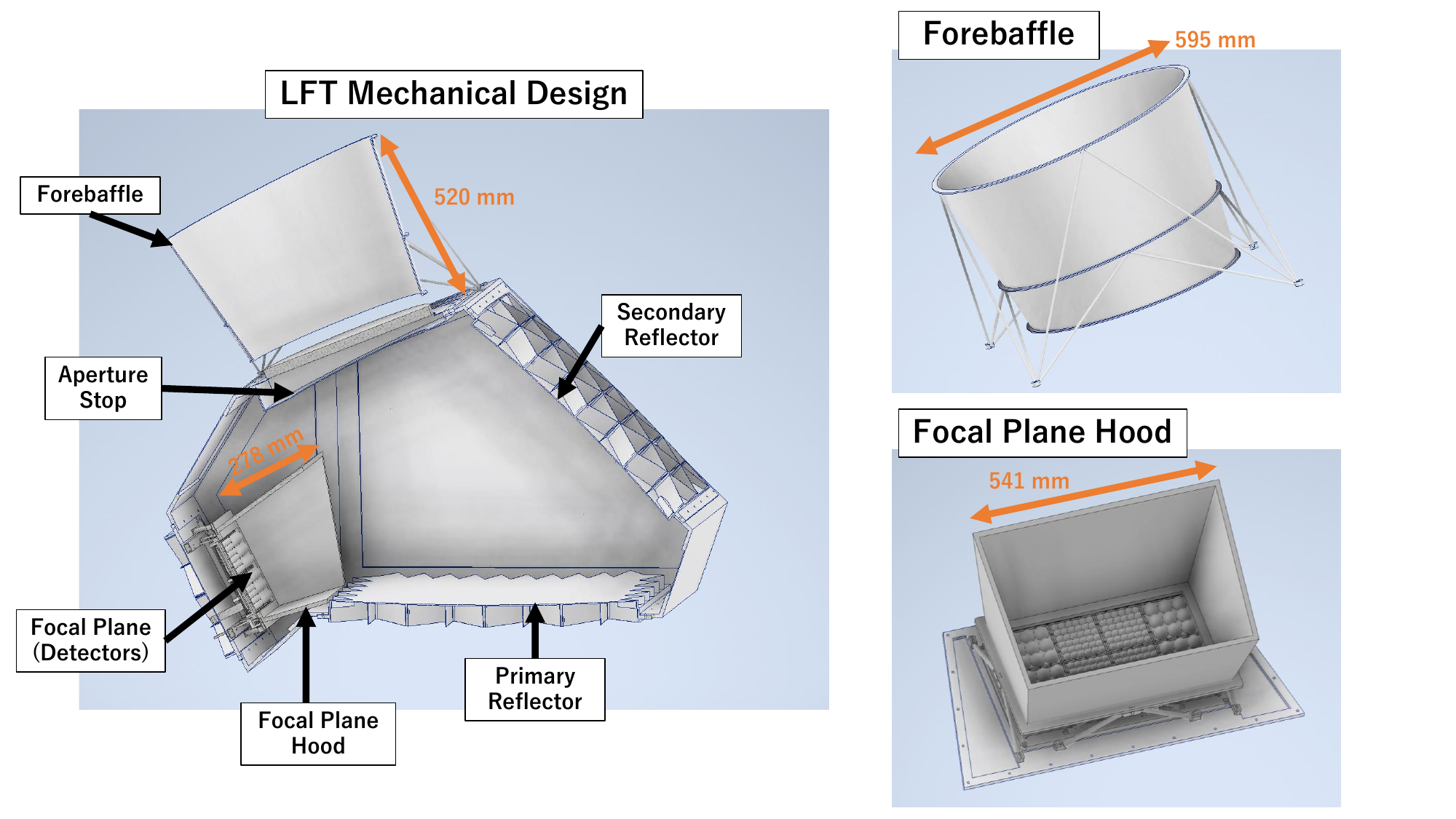}
    \caption{LFT mechanical design including forebaffle and focal plane hood structures. The PMU design is not included, and will be located in the space between the aperture stop and forebaffle structures.}
    \label{fig:lft-mechanical}
\end{figure}

The \LB\ LFT is a wide field-of-view (FOV) reflective telescope optimized for high-precision CMB polarization measurements across nine bands in the 34--161\,GHz frequency range. The overall LFT concept design is described in Sekimoto et al., 2020\cite{Sekimoto_2020}. The focal plane will implement multi-chroic transition-edge sensor (TES) bolometric detectors operated at 100\,mK with lenslet-coupled sinuous antennas\cite{Westbrook_2022}. To achieve high sensitivity, the telescope will be cryogenically cooled to temperatures below 5\,K using mechanical refrigerators and V-groove structures to reduce photon noise. The cryogenic configuration and thermal performance of the PLM is described in Odagiri et al., 2022\cite{Odagiri_2022}. The 5\,K components, including the LFT, are thermally isolated from the service module using a carbon fiber reinforced plastic (CFRP) truss structure. 
The LFT mechanical structure and reflectors will be made of aluminum A6061 to assure uniform thermal contraction throughout the telescope. These structures and reflectors will be machined at ambient temperature (300\,K) accounting for the thermal contraction of 0.4\% such that the optical shapes and alignments will be correct at 5\,K.
The LFT mechanical design is optimized to be light-weight while providing sufficient strength to withstand launch conditions\cite{Oguri_2022}. A polarization modulation unit (PMU) with a continuously rotating half-wave plate (HWP)\cite{Sakurai_2020} will be installed near the aperture stop of the LFT for 1/f noise reduction and mitigation of systematics effects. The LFT also implements an absorptive hood surrounding the focal plane and an absorptive forebaffle sky-side of the aperture stop to mitigate spurious sidelobes due to stray light. 
The absorptive baffling (including the structure and candidate absorber materials) are included in the mass and thermal budgets of the LFT.
The optical design and dimensions of the focal plane hood and forebaffle are explained in Section \ref{sec:po-perf}. The LFT mechanical design is shown in Figure \ref{fig:lft-mechanical}.

\section{\LB\ optical requirements}\label{sec:req}

In order to achieve the \LB\ science goals, detailed analyses and budgeting of the statistical and systematic errors of the \LB\ spacecraft have been done based on observation sensitivity calculations\cite{Hasebe_2023} and cosmological simulations. 
We have determined the optical requirements on the telescopes utilizing a requirements flow-down framework and various trade-off studies. The \LB\ configuration and requirements are explained in detail in \LB\ Collaboration, 2022\cite{LB_ptep_2022}.

As reference, we have summarized the determined requirements\cite{Sekimoto_2020} on the LFT optics (beam and sidelobes) in Table \ref{tbl:opt-req}. The LFT optical design's performance must be verified to meet these requirements using optical simulations during the design phase. 
The built LFT will be verified using pre-flight measurements that will be taken while cooling the telescope to 5\,K in a space chamber to match operation conditions in space. The planned optical measurement technique for the beam pattern\cite{Takakura_2019,Takakura_2022,Nakano_2023} and polarization angle\cite{Takakura_2023} have been demonstrated at room temperature using a quarter scaled LFT. Beam measurements at 5\,K using the same quarter scaled LFT are in-progress\cite{Takakura_2024} as a cryogenic demonstration and to study any optical impacts due to cooling.

\begin{table}[htb]
\small
\begin{center}
\caption{Requirements on the beam and sidelobes properties of the LFT. The cross-polarization requirement is a requirement on the peak cross-polarization beam power. The far sidelobe knowledge precision assumes 0.25 deg$^{2}$ map resolution.}
\begin{tabular}{l|l}
\hline 
Optical Property & Requirement \\
\hline
Frequency                           & 34--161\,GHz \\
Beam width (FWHM)                   & $< 30^{\prime}$ (150\,GHz), $80^{\prime}$ (40\,GHz) \\
Beam ellipticity                    & $< 15$\% \\
Cross-polarization                  & $< -20$\,dB \\
Beam and near sidelobe ($\leq 10^{\circ}$) knowledge    & $-30$\,dB precision \\
Far sidelobe ($> 10^{\circ}$) knowledge                 & $-56$\,dB precision \\
Small scale sidelobe features knowledge                 & $-33$\,dB precision \\
\hline
\end{tabular}
\label{tbl:opt-req}
\end{center}
\end{table}

\section{LFT optical design}\label{sec:opt-design}

\subsection{Trade-off Studies}\label{sec:trade-off}

Crossed Dragone (CD) reflective telescopes have been widely used in the CMB field to achieve high-sensitivity and low cross-polarization measurements across small and large angular scales. A CD is a Mizuguchi-Dragone\cite{Dragone_1978,Mizuguchi_1976} design that consists of off-axis paraboloid primary and hyperboloid secondary reflectors in a crossed-over configuration. CD telescopes have been successfully used by various ground-based CMB experiments such as QUIET\cite{Imbriale_2011} and ABS\cite{Kusaka_2018}. A modified crossed Dragone (MCD) design combined with re-imaging refractive optics has been adopted for the Simons Observatory LAT and Fred Young Submillimeter Telescope (FYST)\cite{Parshley_2018} to achieve improved optical performance for small angular scale CMB observations.

Off-axis Gregorian telescopes have also been used for CMB observations from the ground such as in the Atacama Cosmology Telescope\cite{Fowler_2007}, South Pole Telescope\cite{Padin_2008}, POLARBEAR\cite{PB_2014}, and Simons Array\cite{Stebor_2016}, and also in space by the WMAP\cite{Page_2003} and Planck\cite{Tauber_2010} satellites. While off-axis Gregorian telescopes are easier to baffle to reduce stray light, CD telescopes are known to have clear advantages in low aberrations across a wide diffraction-limited field-of-view (DLFOV), low cross-polarization systematic effects, and compactness\cite{Tran_2008}. We determined that a CD approach is more suitable for the \LB\ LFT. Because the entire LFT will by cryogenically cooled to 5\,K, a CD design allows for a smaller cryogenic mass which is more advantageous for a spacecraft mission in which there are strict mass and cryogenic power constraints.

Small aperture refractive telescopes have also been commonly implemented for large angular scale CMB observations from the ground such as in BICEP\cite{Takahashi_2010}, BICEP2/Keck Array\cite{Ade_2015}, BICEP Array\cite{Hui_2018}, and Simons Observatory SATs\cite{Ali_2020}, and is planned to be implemented for the LiteBIRD MHFT\cite{Montier_2020}. 
The advantage of refractive telescopes is that they can be potentially more compact in size with a similarly wide DLFOV. But they require anti-reflection (AR) coatings on the curved surface lenses which can be technologically challenging when extending to very broadband frequency ranges and complex surface shapes. Refractors also have additional dielectric loss through the refractive lens.

After various trade-off studies, the \LB\ LFT has adopted a MCD reflective design specifically optimized for space-based large angular scale CMB polarization observations. It was determined that the MCD reflective design is capable of achieving the required optical performance levels for the \LB\ science goals while minimizing potential development risks for broadband and low frequency AR coatings applicable for the wide LFT observation frequency range. On the other hand, the \LB\ MHFT are able to adopt a refractive design by splitting the observational frequency range into two smaller range telescopes. For the MHFT, the potential AR coating development risks are mitigated due to the split frequency range, thinner coatings for higher frequencies, and smaller overall component sizes.

Previously, the LFT reflective optics had been designed in a CD-like configuration using anamorphic aspherical surfaces (AAS) in Kashima et al. 2018\cite{Kashima_2018}. Later, the various design constraints such as clearances and crossing angle were reassessed, and the optics was updated in Sekimoto et al. 2020\cite{Sekimoto_2020}. The LFT optics have been further upgrading using a MCD approach. In Section \ref{sec:perf}, we compare the performance of the two designs showing the improvements obtained with the MCD.

\subsection{Design Parameters and Optimization}\label{sec:param-optim}

The design optimization for the LFT was performed in order to satisfy the optical design parameters that are derived from the optical requirements and trade-off studies\cite{Sekimoto_2020}. We summarize the LFT optical design parameters in Table \ref{tbl:opt-params}.

\begin{table}[htb]
\begin{center}
\caption{Design parameters of the LFT optical design derived from the optical requirements and trade-off studies. The beam edge illumination requirements are mainly applicable to the lowest band edge frequency at 34\,GHz that is expected to have largest illumination pattern at the aperture stop and on the reflectors.}
\begin{tabular}{l|l}
\hline 
Design Parameter & Specification \\
\hline
Aperture stop diameter & 400\,mm \\
Field-of-view (FOV) & $18^{\circ} \times 9^{\circ}$ \\
F/\# & 3.0 \\
Crossing angle & $90^{\circ}$ \\
Beam edge illumination at the pupil (aperture stop) & $-3$\,dB \\
Beam edge illumination at serrated reflectors & $-7$\,dB \\
\hline
\end{tabular}
\label{tbl:opt-params}
\end{center}
\end{table}

The aperture stop diameter is derived from the beam width requirement. The field-of-view (FOV) size, F/\#, crossing angle, and beam edge illumination at the stop and reflectors are optimized as a balance to meet the sensitivity requirement while also minimizing stray light and diffraction systematic effects of the LFT sidelobes.
An F/3 optical system was chosen mainly to mitigate and minimize the stray light sidelobes that are characteristic of a CD design. While a smaller F/\# system typically allows for a smaller telescope, it was determined that an F/3 system would fit within the allocated telescope volume and have an acceptable focal plane size\cite{Sekimoto_2020}. The crossing angle of $90^{\circ}$ was chosen as a good balance to moderate both the direct sidelobe and diffuse triple reflection sidelobe stray light paths\cite{Sekimoto_2020}.
To meet the reflector beam edge illumination specification, the geometric optics design optimization was performed under the constraint that 640\,mm diameter (400\,mm main beam + 120\,mm radius margin) collimated rays from the sky shall not be vignetted by the reflectors\cite{imada2018}. 
The clearance between the geometric rays and primary reflector edge near the focal plane is $>30$ mm as margin for diffraction effects.
The optics are also required to be telecentric, that is, couple to a flat focal plane detector array. 

\begin{figure}[htb]
    \centering
    \includegraphics[width=0.65\hsize,pagebox=cropbox,clip]{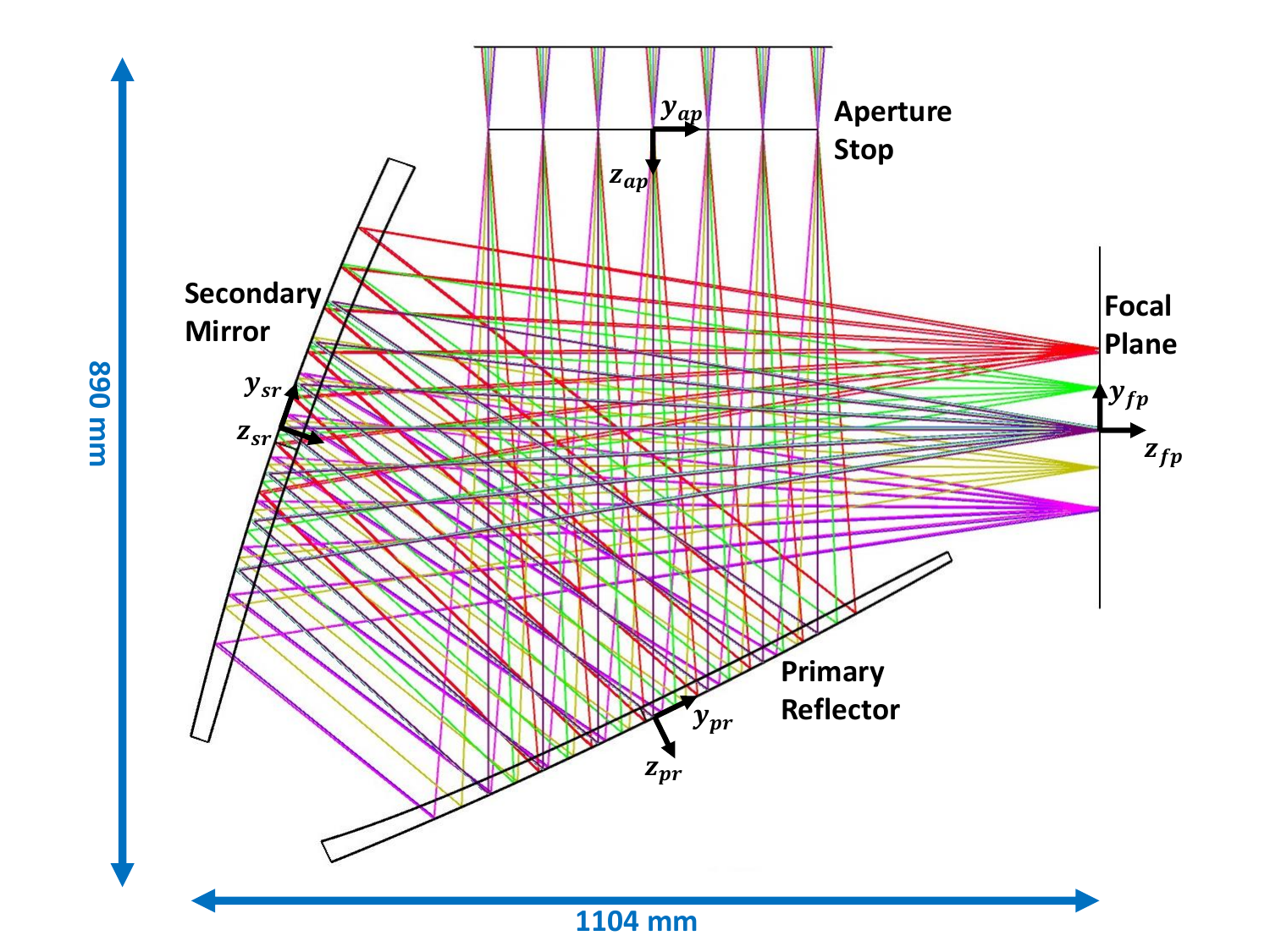}
    \caption{Geometric ray trace of the LFT MCD optics design. The origins of the coordinate systems of the aperture stop $O_\text{ap}$, primary reflector $O_\text{pr}$, secondary reflector $O_\text{sr}$, and focal plane $O_\text{fp}$ are shown. The LFT optics design fits within a 1104\,mm (L) $\times$ 872\,mm (W) $\times$ 890\,mm (H) volume.}
    \label{fig:mcd-go-design}
\end{figure}

The resulting optimized LFT MCD optics consist of a paraboloidal primary reflector and a hyperboloidal secondary reflector both with up to 7th order polynomial correction terms, as will be discussed in the following section. 
The LFT is an F/3 system with a 400\,mm diameter entrance pupil (aperture stop) and a $18^{\circ} \times 9^{\circ}$ DLFOV across its entire observation frequency range. The LFT optical design is shown in Figure \ref{fig:mcd-go-design}. The primary and secondary reflectors are rectangular ($\sim900$ mm $\times$ $\sim800$ mm) to satisfy the beam edge illumination specification. Furthermore, we implement serration rim structures at the reflector edges to reduce diffraction sidelobes. 

\subsection{Modified crossed Dragone design}\label{sec:mcd}

In an MCD design, one modifies a standard CD deisgn with higher order polynomial correction terms added to both the primary and secondary reflectors to correct for the coma aberration and increase the DLFOV. Polynomial correction terms are added relative to the aperture center of each off-axis reflector. A detailed explanation of the adopted design methodology is described in Parshley et al. 2018\cite{Parshley_2018}. One side-effect is that the field curvature is known to increase compared to the original CD design. In ground-based CMB experiment applications, added re-imaging optics have been used to mitigate this non-telecentricity.

For the LFT, we adopted a similar MCD optimization approach. But the wide 34--161\,GHz frequency range and satellite mass requirements, prevented us in implementing re-imaging optics with broadband AR coatings.  
Hence the LFT MCD design was optimized to minimize the field curvature and asymmetries in the F/\# across the focal plane while maximizing the DLFOV.
Specific optimization parameters were added to the optimization merit function to minimize the telecentricity to a flat focal plane, field distortions, and asymmetries in the F/\# in the X and Y directions. While a wide DLFOV size was obtained, this optimization resulted in the peak of the Strehl ratio not being at the center of the FOV but slightly offset in the -Y direction.

In an MCD design, the reflector shapes (sags) are represented as a polynomial expansion with the origin at the center of the reflector aperture. The off-axis paraboloid and hyperboloid reflector shapes of the original CD design are redefined in this new coordinate system, and then the polynomial correction terms are added through optimization. In effect, this introduces on-axis like corrections into an off-axis optical system. The reflector surfaces are defined by 
\begin{equation}
z = \sum_{i=0}^{7} \sum_{j=0}^{7-i} A_{i,j} \left( \frac{x}{R_{N}} \right)^{i} \left( \frac{y}{R_{N}} \right)^{j}
\end{equation}
where $z$ is the surface sag, $A_{i,j}$ is the polynomial coefficient, and $R_{N}$ is a normalization radius. For the LFT, both reflectors have up to 7th order correction terms and $R_{N}=400$ mm. The origins of the reflectors are given in Table \ref{tbl:mcd-coords}. The $A_{i,j}$ values for the LFT primary and secondary reflectors are given in Tables \ref{tbl:mcd-primary} and \ref{tbl:mcd-secondary}. The maximum depth of the primary and secondary reflector surface shapes relative to the reflector edges are $42.3$ mm and $52.4$ mm, respectively. A half-scale secondary reflector has already been fabricated and cryogenically tested as a demonstration\cite{Oguri_2024}.

\begin{table}[htb]
\begin{center}
\caption{Origins of the reflectors and focal plane relative to the aperture stop. The locations of the various coordinate systems are shown in Figure \ref{fig:mcd-go-design}. $\theta_{x}$ is a rotation around the $x$ axis.}
\begin{tabular}{c|cc}
\hline 
Coordinate system & Offset $(y_\text{ap},z_\text{ap})$ [mm] & Rotation $\theta_{x}$ [$^{\circ}$] \\
\hline
Primary reflector $O_\text{pr}$ & $(0,714.19467)$ & $-26.05$ \\
Secondary reflector $O_\text{sr}$ & $(-452.53972,361.90210)$ & $-71.05$ \\
Focal plane $O_\text{fp}$ & $(542.71608,361.90210)$ & $-90.00$ \\
\hline
\end{tabular}
\label{tbl:mcd-coords}
\end{center}
\end{table}

\begin{table*}[t]
\scriptsize
\begin{center}
\caption{Coefficients for the LFT primary reflector.}
\begin{tabular}{|c|c|c|c|c|c|c|c|c|}
\hline 
$A_{i,j}$ & $j=0$ & $j=1$ & $j=2$ & $j=3$ & $j=4$ & $j=5$ & $j=6$ & $j=7$ \\
\hline
$i=0$ & 0 & 0 & -11.739636 & 0.281607 & 0.052087 & -0.118747 & -0.319650 & 0.167694 \\
\hline
$i=1$ & 0 & 0 & 0 & 0 & 0 & 0 & 0 & 0 \\
\hline
$i=2$ & -14.632708 & 0.494182 & 0.427785 & -1.509276 & -0.938836 & 3.316262 & 0 & 0 \\
\hline
$i=3$ & 0 & 0 & 0 & 0 & 0 & 0 & 0 & 0 \\
\hline
$i=4$ & 0.472684 & -1.405746 & -0.447281 & 1.765232 & 0 & 0 & 0 & 0 \\
\hline
$i=5$ & 0 & 0 & 0 & 0 & 0 & 0 & 0 & 0 \\
\hline
$i=6$ & -0.627605 & 2.103679 & 0 & 0 & 0 & 0 & 0 & 0 \\
\hline
$i=7$ & 0 & 0 & 0 & 0 & 0 & 0 & 0 & 0 \\
\hline
\end{tabular}
\label{tbl:mcd-primary}
\end{center}
\end{table*}

\begin{table*}[t]
\scriptsize
\begin{center}
\caption{Coefficients for the LFT secondary reflector.}
\begin{tabular}{|c|c|c|c|c|c|c|c|c|}
\hline 
$A_{i,j}$ & $j=0$ & $j=1$ & $j=2$ & $j=3$ & $j=4$ & $j=5$ & $j=6$ & $j=7$ \\
\hline
$i=0$ & 0 & 0 & 22.625310 & 1.868542 & 0.772344 & 0.209218 & -1.098856 & -0.086693 \\
\hline
$i=1$ & 0 & 0 & 0 & 0 & 0 & 0 & 0 & 0 \\
\hline
$i=2$ & 25.222968 & 2.150801 & 1.009260 & -0.685349 & -1.050082 & 2.488280 & 0 & 0 \\
\hline
$i=3$ & 0 & 0 & 0 & 0 & 0 & 0 & 0 & 0 \\
\hline
$i=4$ & 1.263059 & -0.777701 & -0.326921 & 0.792248 & 0 & 0 & 0 & 0 \\
\hline
$i=5$ & 0 & 0 & 0 & 0 & 0 & 0 & 0 & 0 \\
\hline
$i=6$ & -1.509477 & 1.231467 & 0 & 0 & 0 & 0 & 0 & 0 \\
\hline
$i=7$ & 0 & 0 & 0 & 0 & 0 & 0 & 0 & 0 \\
\hline
\end{tabular}
\label{tbl:mcd-secondary}
\end{center}
\end{table*}

\subsection{Optical performance}\label{sec:perf}

\begin{figure}[!htb]
    \centering
    \includegraphics[width=0.63\hsize,pagebox=cropbox,clip]{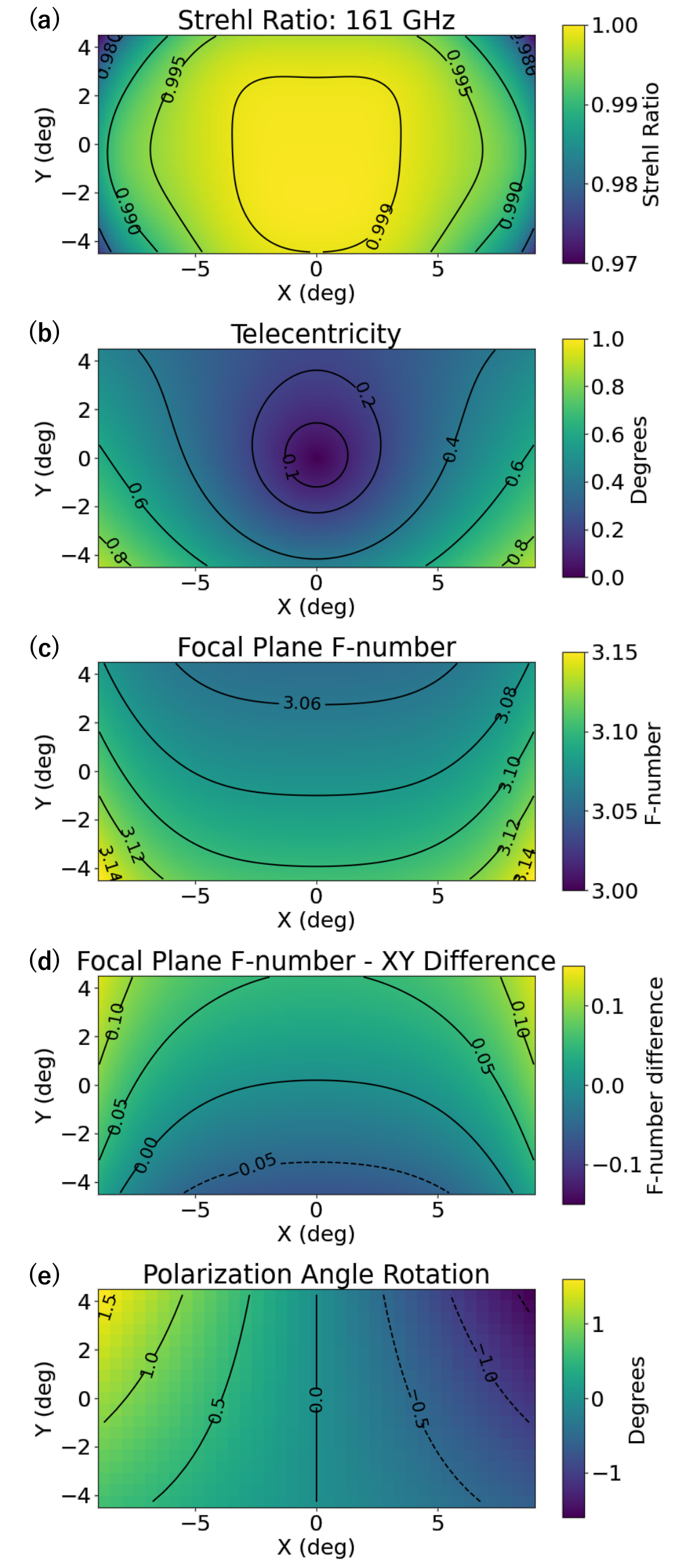}
    \caption{The Strehl ratio at 161 GHz (a), the telecentricity (b), the F-number (c), the F-number difference in X and Y directions (d), and the polarization angle rotation (e) across the FOV of the LFT optical design.}
    \label{fig:opt-performance}
\end{figure}

The LFT optical performance has been measured and optimized using forward time geometric optics (GO) simulations in Zemax OpticStudio\cite{zemax}. The LFT optical design has $18^{\circ} \times 9^{\circ}$ DLFOV with high Strehl ratios $\ge 0.97$ up to 161\,GHz. The optical focal plane area is approximately 420\,mm $\times$ 210\,mm, and is flat with a telecentricity of $\le 0.89^{\circ}$. The average F/\# is 3.09 at the focal plane. In order to minimize asymmetries within the design that can potentially lead to asymmetries in the main beam and sidelobes shapes, we optimized the design to reduce differences in the F/\# in the X and Y directions for each FOV pixel, and on average this F/\# difference is $0.018$. The polarization angle rotation across the FOV is limited to within $\pm1.58^{\circ}$.
The polarization angle rotation was calculated using the polarization ray trace feature in Zemax. The geometric rays in each FOV pixel beam were equally sampled (approximately 2000 rays per beam) and traced from the sky to the focal plane. The calculated polarization angle rotation at the focal plane for each ray was averaged across each pixel beam. This calculation was repeated for the X and Y polarization cases across the FOV.
We summarize the optical performance in Table \ref{tbl:opt-perf}. We also show contour maps of various performance metrics across the FOV in Figure \ref{fig:opt-performance}.

For comparison, the optical performance metrics of the previous AAS LFT design\cite{Sekimoto_2020} are also shown in Table \ref{tbl:opt-perf}. The MCD design out-performs the AAS design in Strehl ratio, telecentricity, F/\# symmetry, and polarization angle rotation across the FOV. The most noticeable difference is in F/\# asymmetry. The AAS design had significant F/\# difference in the X and Y directions per pixel of $0.154$ on average which is reduced to $0.018$ in the MCD design.

We note that, while the requirement on the polarization angle knowledge is strict\cite{LB_ptep_2022,Sekimoto_2020}, the acceptable range of the polarization angle rotation across the FOV is not strict as long as the polarization angles can be accurately calibrated across the FOV. The polarization angle measurement (calibration) technique has been demonstrated to achieve uncertainties of less than $1.9^{\prime}$ using a quarter scaled LFT model at room temperature\cite{Takakura_2023}. Based on this calibration accuracy, the acceptable polarization angle rotation range can be relaxed, and we have determined that the polarization angle rotation across the FOV for the MCD optical design is acceptable.

\begin{table}[htb]
    \begin{center}        
    \caption{Summary of the LFT optical design performance. For each optical parameter the range and average across all FOV (focal plane) pixels are shown. We calculate the F/\# for each pixel by taking the average of the F/\# in the X and Y directions. F/\# asymmetry for each pixel represents the difference between the F/\# in the X direction and Y direction. The average value of the polarization angle rotation within the FOV is zero due to the design symmetry.}
    \begin{tabular}{cc|c|c} 
    \hline
    Optical Parameter &  & MCD & AAS \\
    \hline
    Strehl ratio (161 GHz)  & RANGE     & $\ge$ 0.97 & $\ge$ 0.96 \\
                            & AVG       & 0.99 & 0.99 \\
    \hline
    Telecentricity          & RANGE     & $\le 0.89^{\circ}$ & $\le 0.89^{\circ}$ \\
                            & AVG       & $0.36^{\circ}$ & $0.39^{\circ}$ \\
    \hline
    F/\#, $(f_x+f_y)/2$             & RANGE     & 3.05 $\sim$ 3.15 & 3.01 $\sim$ 3.07 \\
                                    & AVG       & 3.09 & 3.04 \\
    F/\# asymmetry, $|f_x-f_y|$     & RANGE     & $\le$ 0.14 & $\le$ 0.24 \\
                                    & AVG       & 0.02 & 0.15 \\
    \hline
    Polarization angle rotation     & RANGE     & $\pm1.58^{\circ}$ & $\pm1.60^{\circ}$ \\
    \hline
    \end{tabular}
    \label{tbl:opt-perf}
    \end{center}
\end{table}

As is characteristic of a CD design, that the MCD design is originally based on, the LFT MCD design is compact and the main optical components (stop, reflector surfaces, and focal plane surface) are contained within 1104\,mm (L) $\times$ 872\,mm (W) $\times$ 890\,mm (H) volume. The LFT MCD design inherits many of the strengths of a standard CD design such as low telecentricity, high F/\# symmetry, low polarization angle rotation, and compactness while noticeably improving on the Strehl ratio across its wide FOV. The FOV of the LFT MCD design is mainly limited by the reflector beam edge illumination requirement. Thus if this specification were to be relaxed, a larger FOV would be possible because the optical design has sufficiently high Strehl ratio and low telecentricity performance levels beyond the $18^{\circ} \times 9^{\circ}$ FOV.

\subsection{Cross-polarization}\label{sec:xpol}

Using physical optics (PO) simulations, we have analyzed the LFT cross-polarization properties at 34\,GHz across the entire FOV. The optical simulations are constructed and calculated using TICRA Tools\cite{ticratools}. The lowest observation frequency was chosen because, typically, the lowest frequency beams will be more prone to cross-polarization due to having the largest illumination pattern on the reflectors as well as being influenced more by reflector edge diffraction effects\cite{Rodriguez_2013}. The cross-polarization simulation is calculated in reverse time with a symmetric (circular) Gaussian beam feed pattern used as input at the flat focal plane. We used a Gaussian beam feed instead of a realistic lenslet-coupled sinuous antenna detector beam pattern to isolate characterization of the cross-polarization response of the MCD optics without introducing cross-polarization effects due to the detector's antenna design. But we carefully chose the feed's Gaussian beam size to match the realistic detector beam size at 34\,GHz. The reflectors are modeled as perfectly conducting surfaces. The reflector apertures are rectangular with no serrations introduced. 

In CMB polarization experiments, the incoming sky signal as well as the telescope and detector electrical properties can be characterized using the Stokes I, Q, U, and V parameters. I represents unpolarized, Q and U represent the linear polarized, and V represents circularly polarized signal components. Any non-idealities of the telescope optics can create systematic effects that mix the incoming signal's Stokes components, and these effects can be expressed and quantified using a Mueller matrix formalism\cite{odea2007}. The Mueller matrix QU and UQ off-diagonal terms represent the degree of cross-polarization, which mixes the incoming Q and U linear polarization components, in a telescope system. The Mueller QU and UQ terms represent U-to-Q and Q-to-U signal leakage, respectively. These two off-diagonal terms are typically symmetric with a similar power level. In this study, we will mainly focus on the linear polarization components because the CMB $B$ modes of interest are linear polarization signals.

\begin{figure}[htb]
    \centering
    \includegraphics[width=0.9\hsize,pagebox=cropbox,clip]{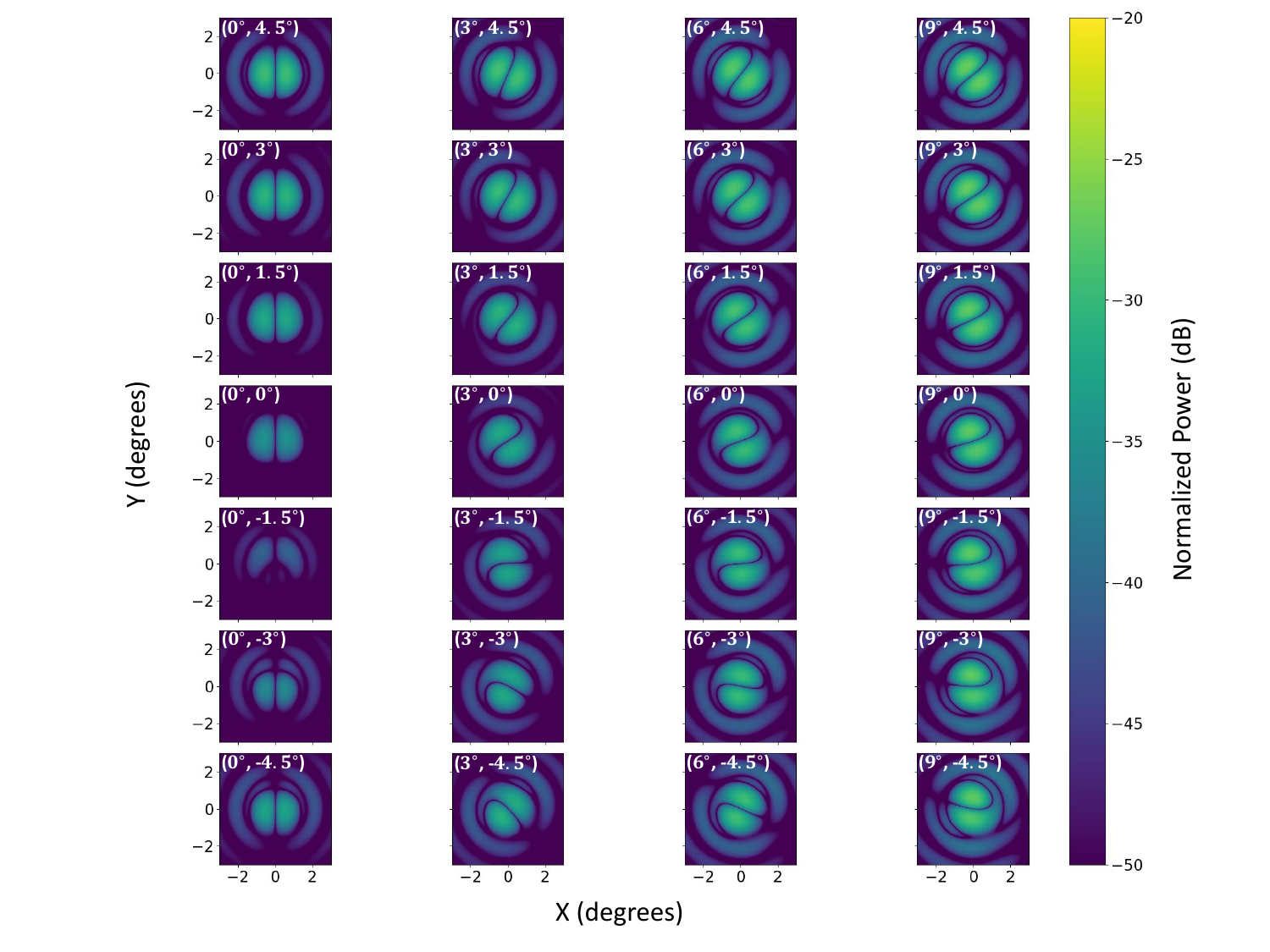}
    \caption{The Mueller QU cross-polarization beam at 34\,GHz across the FOV of the LFT optical design. We calculated the Mueller QU beams for the FOV in the range of $0^{\circ} \le x \le +9^{\circ}$ in increments of $3^{\circ}$, and $-4.5^{\circ} \le y \le +4.5^{\circ}$ in increments of $1.5^{\circ}$. The FOV beam center locations are labeled in white at the top left of each beam map. Because the LFT optics are symmetric along the X direction, the beams in the $-9^{\circ} \le x < 0^{\circ}$ FOV range will have the same response. The cross-polarization beam power is normalized to the peak value of the Mueller II beam.}
    \label{fig:muellerqu}
\end{figure}

Because a telescope's optical sky response is characterized by a beam, the Mueller matrix components will also have a beam-shape response and are typically called the Mueller beams. If the polarization angle is correctly calibrated, the co-polarization beams (QQ, UU terms) will have a monopole (main beam) shaped, and the cross-polarization (QU, UQ terms) will typically be dipole or quadrupole in shaped.
We show the calculated Mueller QU cross-polarization beams in Figure \ref{fig:muellerqu}. The cross-polarization beams are calculated in $3.0^{\circ}$ and $1.5^{\circ}$ increments in the $x$ and $y$ FOV directions, respectively.

For the Mueller QU beams that are off the symmetry plane of the LFT optics, a monopole cross-polarization term will be observed due to the polarization angle rotation across the FOV. In the mission data analysis, this monopole term will be removed by the polarization angle calibration. Thus, in this analysis, the monopole term has been analytically calibrated out\cite{Matsuda_2018}. By symmetry the Mueller UQ beams have a similar pattern and power level.

The LFT optical design has a Mueller QU cross-polarization response of $\le -26.9$ dB across the FOV at 34\,GHz. As expected the cross-polarization response is highest at the top edge corner of the FOV. While the most symmetric dipole response is seen at the center of the FOV, the cross-polarization peak power is smallest not at the center but closer to $(x,y)=(0^{\circ},-1.5^{\circ})$ where the response is $-40.3$ dB. 

Because the Mueller QU (UQ) beams depend on the polarization response difference between X and Y ($45^{\circ}$ and $135^{\circ}$) polarizations, we expect that F\# difference in X and Y directions per pixel should impact the Mueller QU and UQ beams as well. It can be seen that the dipole direction has a tendency to follow the contour lines of the F\# difference plot in Figure \ref{fig:opt-performance}.
Based on this cross-polarization characterization at 34\,GHz, we expect that the LFT optical design satisfies the cross-polarization requirement and will be $\le -26.9$ dB across the entire FOV.

\section{Main beam and sidelobes performance}\label{sec:po-perf}

\begin{figure}[htb]
    \centering
    \includegraphics[width=0.9\hsize,pagebox=cropbox,clip]{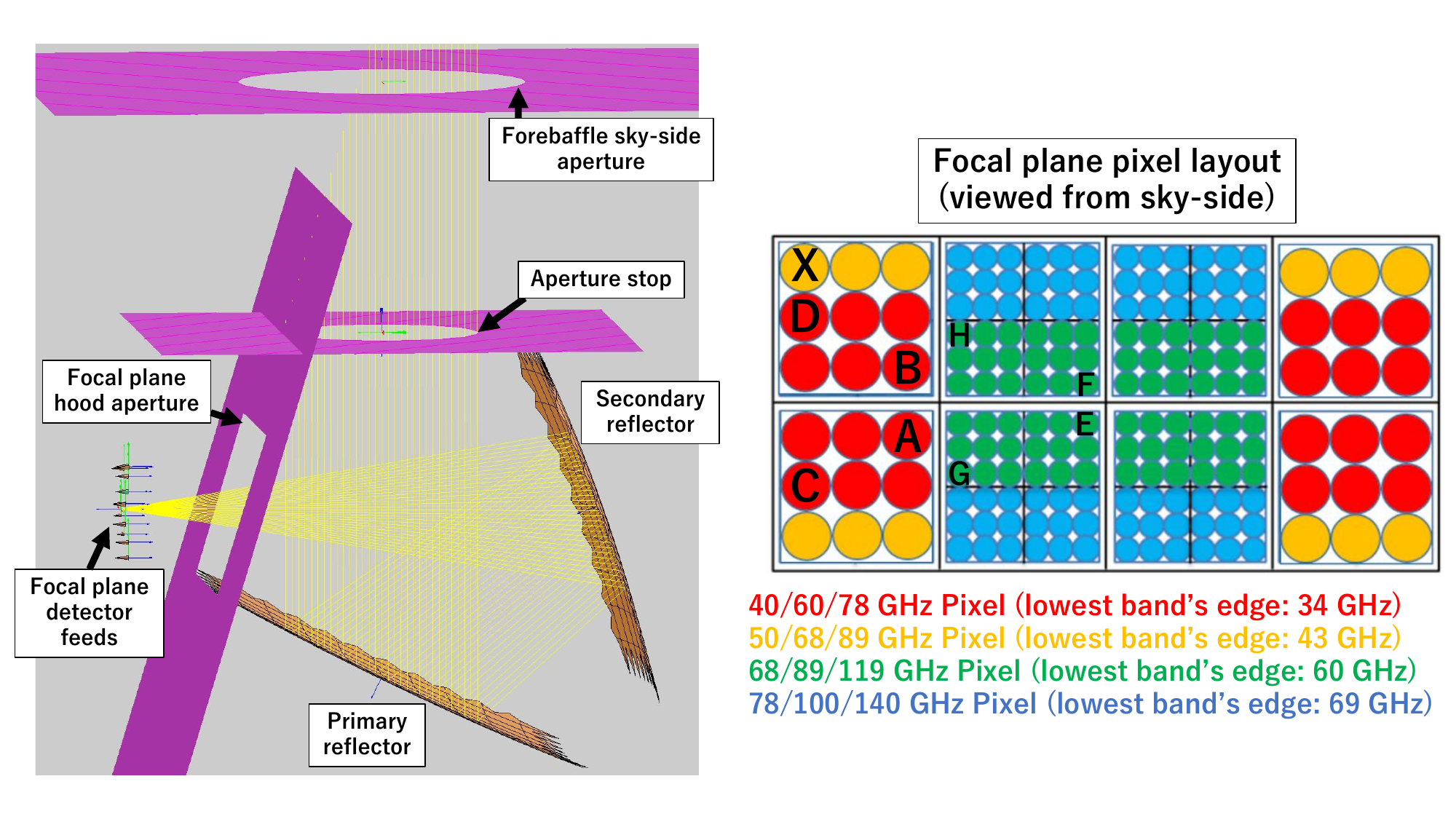}
    \caption{Optics (PO/MoM) model of the LFT (left) and focal plane pixel layout (right). The tri-chroic pixel types are labeled in different colors, and we also indicate the lowest band's edge frequency for each pixel type.}
    \label{fig:ticra-model}
\end{figure}

We have used PO and method of moments (MoM) simulations to assess the LFT main beams and sidelobe levels including diffraction effects at specific frequencies. The LFT implements tri-chroic lenslet-coupled sinuous antenna detector pixels\cite{Sobrin_2022} in its focal plane as shown in Figure \ref{fig:ticra-model}. There are four different types of pixels in the focal plane, and each pixel is capable of observing at three different frequency bands simultaneously. Two types of pixels have 32\,mm diameter lenslets, and the other two types of pixels have 16\,mm diameter lenslets. For the two lenslet sizes, the lower frequency 40 and 68\,GHz detectors are placed closer to the middle of the FOV compared to the 50 and 78\,GHz detectors to minimize potential sidelobes due to diffraction from the various optical elements.

We expect that the 32\,mm diameter lenslet 40\,GHz and the 16\,mm diameter lenslet 68\,GHz center frequency bands will have the broadest beam radii and the largest spillover (edge taper illumination) power at the aperture stop in the \LB\ observational bands. The optical design is diffraction-limited across all frequencies and the aberration effects are expected to be sub-dominant compared to diffraction effects in the main beam. 
By design, for tri-chroic pixels the edge taper illumination power will decrease as a function of higher frequency and the diffraction effects in the main beam will become less prominent.
Thus, if the lowest frequency beams for each lenslet size satisfy the 
beam width and ellipticity (upper limit) 
requirements, we expect that the beams at higher frequencies, that have lower edge taper illumination, will comply with the optical performance requirements. Therefore, we simulated the beams at 34 and 60\,GHz, which is the lowest bands' edge frequencies in the 40 and 68\,GHz center frequency bands.

We have developed TICRA Tools optical simulation models of the LFT optics. The model consists of perfectly conductive primary and secondary reflectors with serrations, an aperture stop, a perfectly absorptive focal plane hood, and a perfectly absorptive forebaffle as shown in Figure \ref{fig:ticra-model}. The reflector serrations are triangular with $\sim60$ mm height and $\sim70$ mm pitch. The optical impacts of the serrations on the LFT performance have been studied in detail in Matsuda et al., 2024\cite{Matsuda_2024}. 
The focal plane hood is a rectangular cone shaped baffle that extends from the optical focal plane toward the secondary reflector. 

In this simulation model, the rectangular cone extends 278\,mm out perpendicular from the focal plane with a 541\,mm $\times$ 347\,mm rectangular aperture, and is cut diagonally such that the hood extends out further on the top side to prevent stray light from directly entering the detectors from the aperture stop. Thus the hood opening aperture is a trapezoid with bases 541\,mm and 508\,mm and a height of 346\,mm.
The forebaffle is a circular cone shaped baffle that extends 520\,mm out perpendicular from the aperture stop with 595\,mm sky-side diameter.

The focal plane hood and forebaffle models perfectly absorptive to reduce the simulation time. In reality, these baffling structures will implement absorbers with finite absorptivity. The impact of this finite absorptivity on the sidelobe power levels are analytically estimated in this study when comparing to the \LB\ sidelobe requirements. 

There are various potential systematic effects associated with lenslet-coupled sinuous antenna detectors such as the impact of detector cross-polarization, detector sidelobes due to the anti-reflection coated lenslets\cite{Westbrook_2022}, and polarization angle wobble of the sinuous antenna. Potential imperfections in the antennas and lenslets during fabrication can also influence the level of these systematic effects. The systematic effect of the polarization angle wobble as a function of frequency can be mitigated by carefully choosing the orientations of the sinuous antennas such that there are mirror image antennas on each wafer to provide polarization rotation cancellation\cite{Westbrook_2020}. The systematic impact of the detector's cross-polarization, polarization angle wobble, and potential fabrication imperfections on the polarization performance of the LFT across all observational frequencies is an area of future study. The results of this paper will focus on the LFT main beam and sidelobe performances including the impact of the lenslet-coupled sinuous antenna detector's main beam and sidelobe response.

\subsection{Main beam}\label{sec:po-main}

To assess the LFT's main beam performance, we simulated the 34\,GHz pixel with 32 mm diameter lenslet and 60\,GHz pixel with 16 mm diameter lenslet because these pixel beams are the broadest and those with the highest edge taper power at the aperture stop. The pixel beam patterns (including detector sidelobes) of the combined sinuous antenna and 32-mm and 16-mm silicon lenslet with a two-layer anti-reflection coating at 34\,GHz and 60\,GHz, respectively, were simulated using HFSS\cite{hfss}, and were used as input into the TICRA Tools model for the 34 and 60\,GHz feeds at the focal plane. Because beam and sidelobe properties are expected to degrade for pixels further away from the center of the focal plane, we have chosen to simulate the 34 and 60\,GHz pixels located at the most inner and outer corners of the respective detector wafers to show the range of the beam characteristics across the focal plane. We show the simulation model and simulated pixel positions on the focal plane in Figure \ref{fig:ticra-model}.

\begin{table*}[t]
\begin{center}
\caption{Simulated main beam properties at 34 and 60\,GHz.}
\begin{tabular}{ccc|cccc}
\hline 
Feed & Frequency & Polarization & FWHM & FWHM & Major Axis & Ellipticity \\
& & Direction & (major) & (minor) & Angle & \\
\hline
Pixel A & 34 GHz & X & $77.33^{\prime}$ & $75.78^{\prime}$ & $69.17^{\circ}$ & $1.01$\% \\
& & Y & $77.01^{\prime}$ & $76.13^{\prime}$ & $171.13^{\circ}$ & $0.57$\% \\
Pixel B & 34 GHz & X & $77.34^{\prime}$ & $75.83^{\prime}$ & $74.15^{\circ}$ & $0.99$\% \\
& & Y & $77.03^{\prime}$ & $76.13^{\prime}$ & $161.53^{\circ}$ & $0.57$\% \\
Pixel C & 34 GHz & X & $77.06^{\prime}$ & $76.50^{\prime}$ & $65.15^{\circ}$ & $0.36$\% \\
& & Y & $77.71^{\prime}$ & $75.88^{\prime}$ & $168.50^{\circ}$ & $1.19$\% \\
Pixel D & 34 GHz & X & $77.67^{\prime}$ & $76.02^{\prime}$ & $56.26^{\circ}$ & $1.07$\% \\
& & Y & $77.63^{\prime}$ & $76.06^{\prime}$ & $5.08^{\circ}$ & $1.02$\% \\
\hline
Pixel E & 60 GHz & X & $42.93^{\prime}$ & $42.85^{\prime}$ & $47.64^{\circ}$ & $0.09$\% \\
& & Y & $42.90^{\prime}$ & $42.88^{\prime}$ & $52.72^{\circ}$ & $0.02$\% \\
Pixel F & 60 GHz & X & $42.91^{\prime}$ & $42.89^{\prime}$ & $82.64^{\circ}$ & $0.03$\% \\
& & Y & $42.93^{\prime}$ & $42.87^{\prime}$ & $125.78^{\circ}$ & $0.07$\% \\
Pixel G & 60 GHz & X & $43.16^{\prime}$ & $42.87^{\prime}$ & $166.88^{\circ}$ & $0.33$\% \\
& & Y & $43.19^{\prime}$ & $42.89^{\prime}$ & $162.25^{\circ}$ & $0.39$\% \\
Pixel H & 60 GHz & X & $43.23^{\prime}$ & $42.83^{\prime}$ & $30.57^{\circ}$ & $0.46$\% \\
& & Y & $43.21^{\prime}$ & $42.87^{\prime}$ & $27.81^{\circ}$ & $0.39$\% \\
\hline
\end{tabular}
\label{tbl:opt-beam}
\end{center}
\end{table*}

\begin{figure}[htb]
    \centering
    \includegraphics[width=0.7\hsize,pagebox=cropbox,clip]{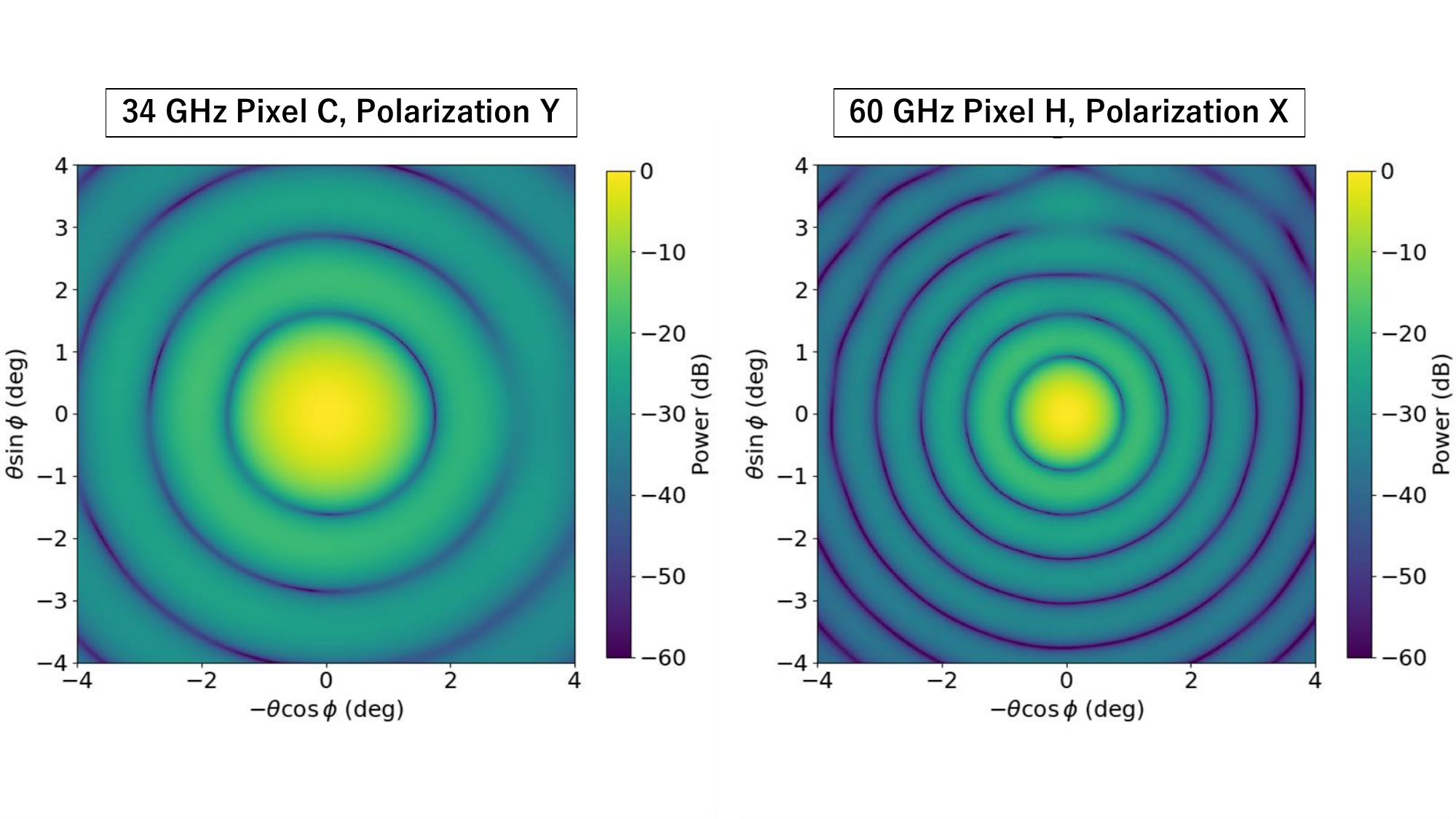}
    \caption{Simulated main beams for 34 GHz pixel C in Y polarization (left) and 60 GHz pixel H in X polarization (right).}
    \label{fig:opt-mainbeam}
\end{figure}

We summarize the main beam properties of the simulated pixels in Table \ref{tbl:opt-beam}. Beams for X and Y polarization detectors are simulated for each pixel position. The simulation results show that the LFT optical design satisfies the beam requirements with beams sizes $<78^{\prime}$ at the lowest 34\,GHz and beam ellpticities $<1.2\%$. In general, the beam ellipticies are small across the focal plane with the beams being more symmetric near the center of the focal plane as expected. The beams are less elliptical for the 60\,GHz pixels compared to the 34\,GHz pixels.
Example simulated main beams are shown in Figure \ref{fig:opt-mainbeam} for the 34 and 60\,GHz pixels with the largest ellipticity.

As an illustrative exercise and check, we also simulated a hypothetical 34\, GHz pixel placed at the corners of the focal plane and found that the ellipticities are $<1.7\%$. This gives us more confidence that the beam ellipticities should be small across the focal plane and meet the requirements.
It should be noted that, in each tri-chroic pixel, the resulting LFT beam ellipticity at higher frequencies should depend more on any innate detector beam ellipticity because the edge taper illumination at the aperture stop is smaller. More detailed optical simulations of the main beams for all frequencies across the focal plane will be assessed in a future analysis.

It was found that the absorptive forebaffle slightly increases the main beam ellpiticity for the lowest 34\,GHz pixels. It is hypothesized that the diffraction effects off the forebaffle aperture, which are most dominant at the lowest frequencies, are adding a non-negligible background and affecting the beam shape to a small degree. This impact is small and the beam ellipticities satisfy the requirements with the absorptive forebaffle. 
The ellipticity differences between X and Y polarization detector pairs are small and not expected to be problematic. Furthermore, the continuous modulation by the HWP can mitigate this systematic effect.

\subsection{Sidelobes}\label{sec:po-sidelobes}

\begin{figure}[htb]
    \centering
    \includegraphics[width=0.65\hsize,pagebox=cropbox,clip]{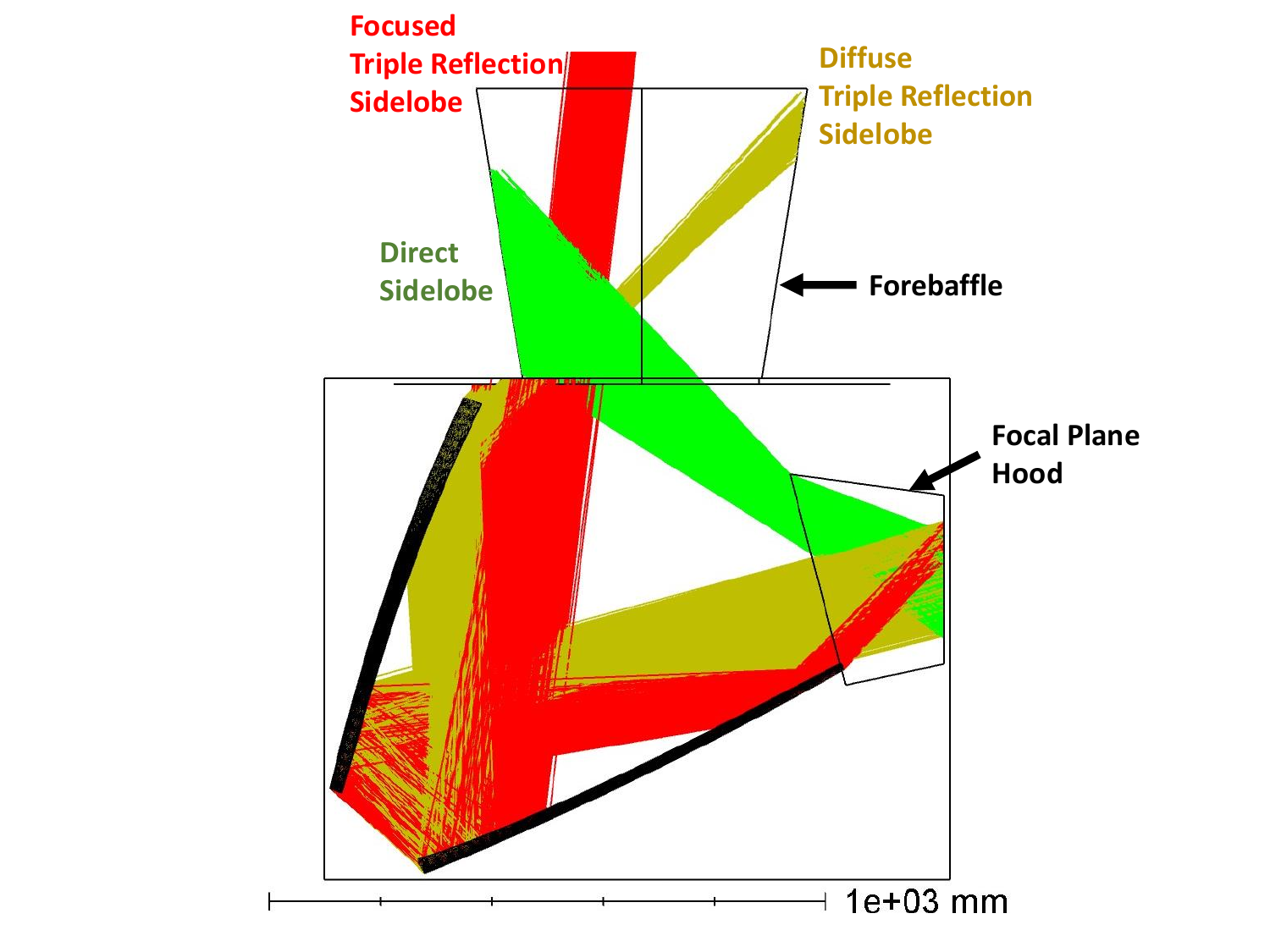}
    \caption{LFT sidelobe ray path simulations using geometric optics. The focal plane hood and forebaffle effectively block the direct sidelobe and diffuse triple reflection sidelobe.}
    \label{fig:sidelobe-go}
\end{figure}

From GO simulations, we expect that the LFT will have stray light paths that are characteristics of a standard CD design. These multiple reflection sidelobe paths are shown in Figure \ref{fig:sidelobe-go}. In this study, we will focus on estimating the resulting LFT sidelobe power due to the direct sidelobe, diffuse triple reflection sidelobe, and the focused triple reflection sidelobe, explained below. Other stray light sidelobe paths can exist, but other sidelobe paths will reflect or scatter at least once off the LFT mechanical inner walls lined with absorbers. Thus these other paths will depend on the detailed LFT mechanical design and is not necessarily a characteristic of a CD optics design. The other sidelobe paths are out of the scope of this paper and will be assessed when the detailed LFT mechanical design and absorber material developments progress further.

The direct sidelobe is a stray light path in which light from the sky can directly enter the detectors without reflecting off the reflectors. The diffuse triple reflection sidelobe is a stray light path in which light from the sky reflects three times off the reflectors in the order of secondary, primary, and secondary once again before enter the detectors. Similarly, the focused triple reflection sidelobe also reflects three times off the reflectors but in the order of primary, secondary, and primary once again before entering the detectors. The direct sidelobe and diffuse triple reflection sidelobe can theoretically be mitigated using a sufficiently long forebaffle.

\begin{figure}[htb]
    \centering
    \includegraphics[width=0.9\hsize,pagebox=cropbox,clip]{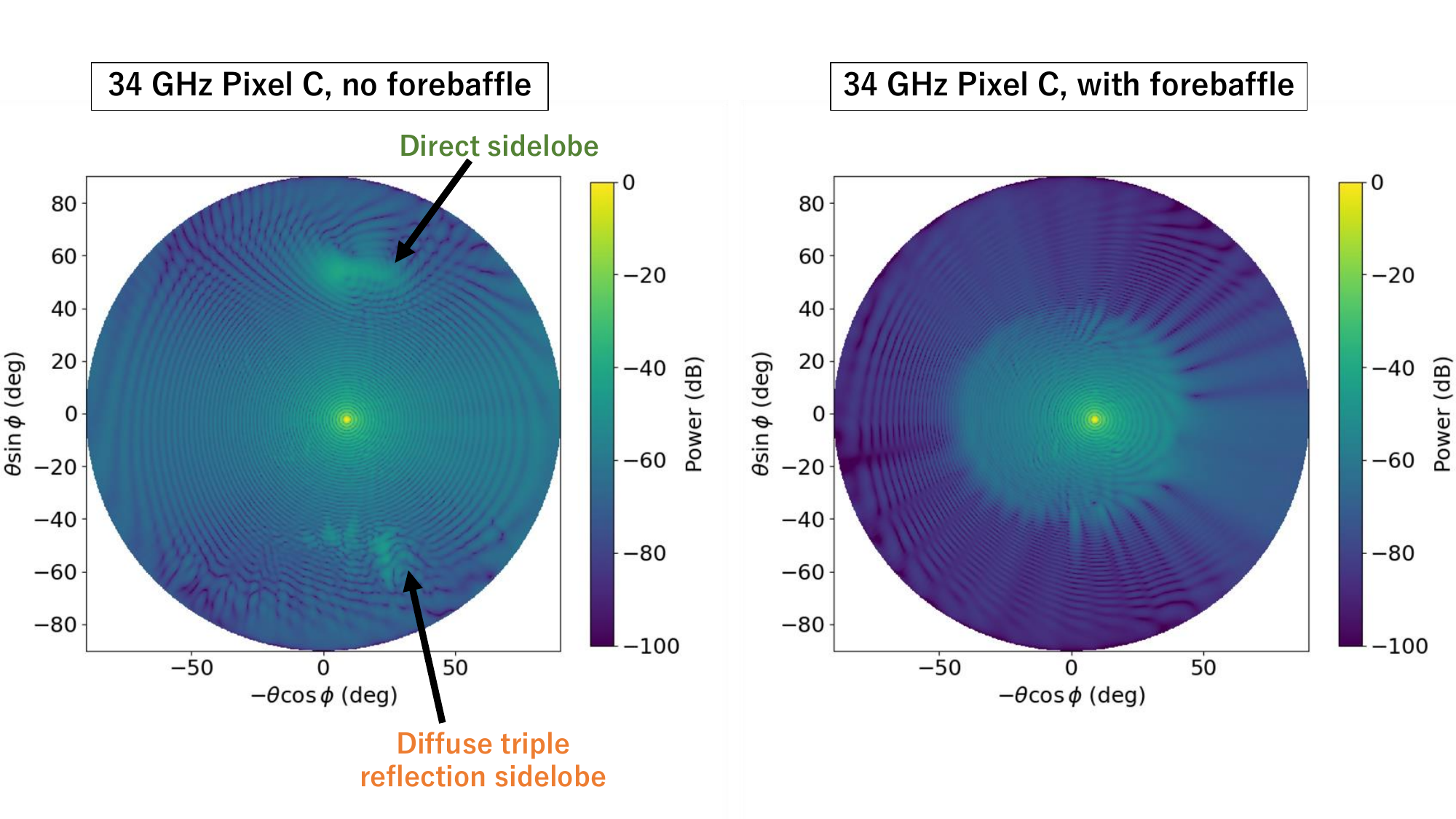}
    \caption{LFT sidelobe simulation results for 34\,GHz pixel C without a forebaffle (left) and with an absorptive forebaffle (right).}
    \label{fig:sidelobe-pixc}
\end{figure}

\begin{figure}[htb]
    \centering
    \includegraphics[width=0.9\hsize,pagebox=cropbox,clip]{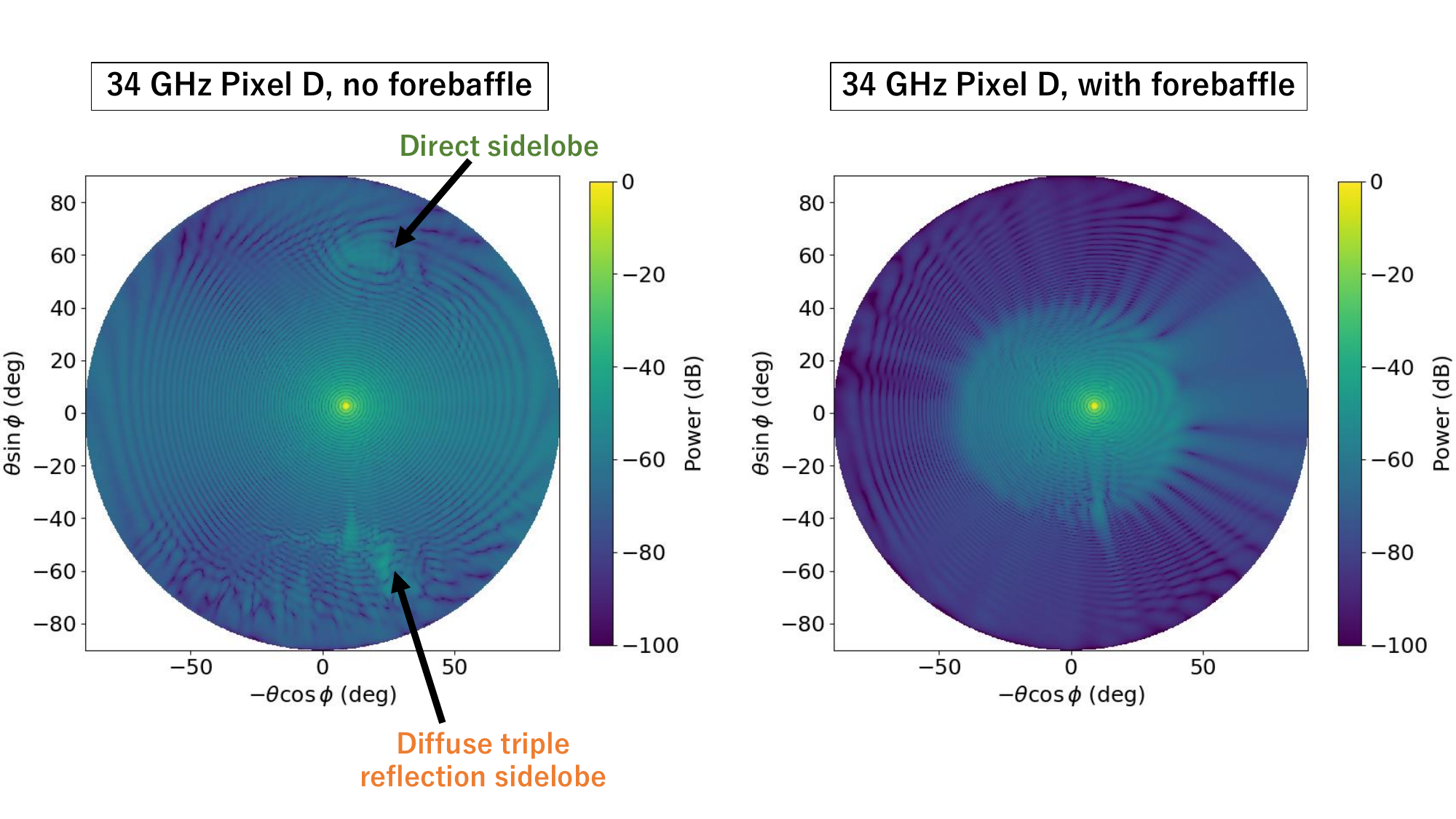}
    \caption{LFT sidelobe simulation results for 34\,GHz pixel D without a forebaffle (left) and with an absorptive forebaffle (right).}
    \label{fig:sidelobe-pixd}
\end{figure}

The LFT sidelobes are expected to be the greatest for the lowest observation frequencies and the edge pixels of the focal plane. The lowest frequencies will have the largest beams and diffraction effects, and the edge pixels will interact with the reflector edges the strongest to create sidelobe effects. Therefore, we simulated the sidelobes for the outer-most edge 34\,GHz pixels, positions C and D on the focal plane, to give an upper limit on the relevant sidelobe power. We show the sidelobe simulation results for pixels C and D in Figures \ref{fig:sidelobe-pixc} and \ref{fig:sidelobe-pixd}, respectively. For each case, we show simulation results for with and without an absorptive forebaffle.

We found that the direct and diffuse triple reflection sidelobes are observable at 34\,GHz only for simulations without a forebaffle. The direct sidelobe is most prominent in the pixel C and has a peak power of -40.5 dB and -54.0 dB relative to the main beam peak for pixels C and D, respectively. This sidelobe power is expected to be larger for focal plane pixels located on the lower half of the focal plane as observable from the GO simulation results. The diffuse triple reflection sidelobe has a peak power of -45.4 dB and -45.9 dB for pixels C and D, respectively. These sidelobe paths and angular far field locations match the expectations based on GO simulations.

As expected, both the direct sidelobe and diffuse triple reflection sidelobe can be mitigated using an absorptive forebaffle. In the case of a perfectly absorptive forebaffle, these sidelobe features are mitigated, and we can only observe sidelobes created from diffraction. The cone-shaped forebaffle is designed to mitigate sidelobes that extend approximately beyond $\gtrsim 40^{\circ}$ from boresight as can be seen from the simulation results.
While geometric optics show that these sidelobes can be fully mitigated by the current forebaffle size, in pixel D a remnant sidelobe feature is still apparent. We hypothesize that this is due to the diffuse triple reflection sidelobe being absorbed (truncated) very close to the forebaffle aperture and leaving a small diffraction feature. We expect that extending the forebaffle length will help in further mitigating this sidelobe power.

In reality, the absorptivity of the forebaffle will be finite, and candidate absorber materials have a reflectivities of approximately $\le-17$ dB depending on the material. 
With this assumption, we expect that the peak sidelobe power at the lowest observation frequency can be reduced to approximately $<-57$ dB which is below the required far sidelobe knowledge level.

The LFT optics are designed such that the focused triple reflection sidelobe is only possible for pixels located near the top edge portion of the 420\,mm $\times$ 210\,mm focal plane area. Due to mechanical considerations, the focal plane edge pixel centers lie more internal to this focal plane area, and thus we expect that the focused triple reflection sidelobe is small or non-existent for these top edge pixels. Simulations at 34\,GHz confirm that this focused triple reflection sidelobe is not observed. 

To determine an upper limit on this possible sidelobe from the top edge pixels, we ran simulations by placing a hypothetical 34\,GHz pixel at the most outer top corner 42\,GHz pixel location of the focal plane, labeled with an X on Figure \ref{fig:ticra-model}. The focused triple reflection sidelobe was found to have a peak power of -37.0 dB relative to the main beam peak for this hypothetical 34\,GHz pixel. The sidelobe power will theoretically decrease for higher frequencies pixels, and therefore we expect that the focused triple reflection sidelobe peak power is conservatively $<-37$ dB which is below the required small scale sidelobe features knowledge level.

\section{Conclusions}

In this paper we present the performance of the \LB\ LFT optical system based on the modified crossed Dragone design.
We further assess the optical performance of the design using geometric optics, physical optics, and method of moments optical simulations and compare to the \LB\ optical requirements and specifications. The LFT optical design has a $18^{\circ} \times 9^{\circ}$ FOV with Strehl ratios $\ge 0.97$, telecentricity $\le 0.89^{\circ}$, average F/\# of $3.09$, F/\# XY symmetry $\le 0.14$, and polarization angle rotation limited to within $\pm 1.58^{\circ}$ across the entire FOV. The cross-polarization Mueller QU beam response is estimated to be $\le -26.9$ dB. The LFT design inherits many of the strengths of the original CD design of which it has been optimized from while improving on its optical performance. 

We estimated the main beam and sidelobe response of the LFT design including ideal baffling for outermost 34 and 60\,GHz frequency detector pixels to achieve an upper limit on these response levels. The LFT has beam sizes $< 78^{\prime}$ and beam ellipiticies $< 1.2\%$ at 34\,GHz, which satisfy the main beam requirements. We simulated the direct, diffuse triple reflection, and focused triple reflection sidelobes characteristic of CD-like configurations. The direct and diffuse triple reflections sidelobes are expected to be sufficiently mitigated by the focal plane hood and forebaffle to $< -57$ dB, which is below the far sidelobe knowledge requirement. The focused triple reflection sidelobe cannot be attenuated by the baffling, but is expected to be $< -37$ dB, below the small scale sidelobe features sidelobe knowledge requirement.

The MCD optical design of the LFT satisfies the optical design requirements and specifications. Therefore the presented LFT optical design is a suitable design to for high-sensitivity and high-precision CMB polarization measurements to achieve the \LB\ science goals, and is implemented as the \LB\ LFT baseline design. From this study, we found that an MCD design can be an attractive design choice for space-based CMB experiments.

While the basic optical performance levels are assessed in this study to compare to the optical design requirements and specifications, more studies and simulations are needed to characterize the main beam and sidelobes features across the entire observation frequency range. Specifically, we will develop optical simulation models that include detailed detector main beam and sidelobes, telescope mechanical structures, finite absorptivity baffling, and other PLM structures such as the V-grooves and MHFT structures. 
The LFT's beam and sidelobe patterns will be optically measured pre-flight at cryogenic temperatures to verify that the LFT satisfies the optical requirements.

\begin{backmatter}

\bmsection{Acknowledgments}
This work is supported in Japan by ISAS/JAXA for the Pre-Phase-A2 studies, by the acceleration program of the JAXA Research and Development Directorate, by the World Premier International Research Center Initiative (WPI) of MEXT, and by the JSPS Core-to-Core Program. 

\bmsection{Disclosures}
The authors declare no potential conflicts of interest.

\bmsection{Data availability}
Data underlying the results presented in this paper are not publicly available at this time but may be obtained from the authors upon reasonable request and institutional approval.

\end{backmatter}


\bibliography{references}

\end{document}